\documentclass[prd,amsmath,notitlepage,twocolumn]{revtex4-2}
\usepackage{graphicx}
\usepackage{float}
\usepackage{makecell}
\usepackage{flushend}
\usepackage{epstopdf,cancel}
\usepackage{epsf,latexsym,bbm,euscript}
\usepackage{amssymb,amsmath,amsthm}
\usepackage{mathtools} 
\usepackage{times,graphics}
\usepackage{soul,xcolor}
\usepackage{mathtools}
\usepackage{multirow}

\usepackage{url,hyperref}
\hypersetup{colorlinks,linkcolor={blue!55!black},citecolor={red!50!black},urlcolor={blue!45!black},breaklinks=true}

\usepackage{scalerel}
\usepackage{tikz}
\usetikzlibrary{svg.path}
\definecolor{orcidlogocol}{HTML}{A6CE39}

\def\6{{\langle}}
\def\9{{\rangle}}
\newcommand{\defeq}{\vcentcolon=}
\newcommand{\eqdef}{=\vcentcolon}

\newcommand{\ph}{\varphi}
\newcommand{\tht}{\vartheta}
\newcommand{\eps}{\varepsilon}

\newcommand{\tcb}{\textcolor{blue}}			
\newcommand{\tcr}{\textcolor{red}}

\usepackage{scalerel}
\usepackage{tikz}
\usetikzlibrary{svg.path}
\definecolor{orcidlogocol}{HTML}{A6CE39}
\tikzset{
	orcidlogo/.pic={
		\fill[orcidlogocol] svg{M256,128c0,70.7-57.3,128-128,128C57.3,256,0,198.7,0,128C0,57.3,57.3,0,128,0C198.7,0,256,57.3,256,128z};
		\fill[white] svg{M86.3,186.2H70.9V79.1h15.4v48.4V186.2z}
		svg{M108.9,79.1h41.6c39.6,0,57,28.3,57,53.6c0,27.5-21.5,53.6-56.8,53.6h-41.8V79.1z M124.3,172.4h24.5c34.9,0,42.9-26.5,42.9-39.7c0-21.5-13.7-39.7-43.7-39.7h-23.7V172.4z}
		svg{M88.7,56.8c0,5.5-4.5,10.1-10.1,10.1c-5.6,0-10.1-4.6-10.1-10.1c0-5.6,4.5-10.1,10.1-10.1C84.2,46.7,88.7,51.3,88.7,56.8z};
	}
}

\newcommand\orcidlink[1]{\href{https://orcid.org/#1}{\mbox{\scalerel*{
				\begin{tikzpicture}[yscale=-1,transform shape]
					\pic{orcidlogo};
			\end{tikzpicture}}{X}}}}

\newcommand{\be}{\begin{equation}}
\newcommand{\ee}{\end{equation}}
\newcommand{\ba}{\begin{eqnarray}}
\newcommand{\ea}{\end{eqnarray}}

\newcommand{\nn}{\nonumber}

\newcommand{\mS}{{\mathrm{S}}}
 \newcommand{\mA}{{\mathrm{A}}}
 \def\at{\mathsf{t}}

\newcommand{\sccur}{{\mathcal{R}}}

\def\eD{\EuScript{D}}
\def\eK{\EuScript{K}}
\def\eV{\EuScript{V}}

\def\rin{\mathrm{in}}
\def\rout{\mathrm{out}}
\def\rmax{\mathrm{max}}

\def\etal{\textit{et al.}}

\def\bnab{{\mbox{\boldmath{$\nabla$}}}}
\def\anab{{\mbox{\sf\boldmath{{$\nabla$}}}}}
\def\half{{\tfrac{1}{2}}}

\def\pad{{\partial}}

\def\ha{{\hat{a}}}

\usepackage[caption=false]{subfig}
\newcommand{\subfigimg}[3][,]{%
  \setbox1=\hbox{\includegraphics[#1]{#3}}
  \leavevmode\rlap{\usebox1}
  \rlap{\hspace*{-10pt}\raisebox{.5\baselineskip}{\small{#2}}}
  \phantom{\usebox1}
}

\def\sg{\textsl{g}}

\def\eC{\EuScript{C}}

\def\cO{\mathcal{O}}

\def\rA{\mathrm{A}}

\def\mfr{\mathfrak{r}}

\begin{document}

\title{Physical black holes: spacetime and matter near trapping surfaces}

\author{Swayamsiddha Maharana\,\orcidlink{0009-0004-6006-8637}}
\email{swayamsiddha.maharana@hdr.mq.edu.au}

\author {Rama Vadapalli\,\orcidlink{0000-0001-7470-1342}}
\email{venkataramaraju.vadapalli@hdr.mq.edu.au}

\affiliation{School of Mathematical and Physical Sciences, Macquarie University, NSW 2109, Australia}

\begin{abstract}
Within a semiclassical framework, we investigate spherically symmetric solutions of the Einstein equations that (i) develop a trapped region within a finite time as measured by distant observers, and (ii) remain sufficiently regular at the apparent horizon. These requirements characterize the self-consistent approach, and it imposes specific constraints on the limiting form of the energy momentum tensor (EMT). Previous works \cite{Finite_energy:2019,BMMT:2019,MMT:2022,Terno:2020} argued that only certain EMT classes admit viable dynamical solutions, while discarding others (in Ref.~\cite{Terno:2020}) based on reasoning that is not entirely robust. Here, we explicitly demonstrate that the discarded classes are indeed unphysical. In addition, we also introduce a new class of static solutions that are regular at the horizon and allow finite-time horizon crossing according to a distant observer.
\end{abstract}

\maketitle

\section{Introduction}
\noindent Astrophysical black holes (ABHs)~\cite{Frolov:1998,bambi:2017} are dark, massive ultracompact objects (UCOs), characterized by extremely strong gravitational fields, whose existence is firmly supported by numerous observations. Various theoretical models, both with and without horizons, have been proposed to describe such UCOs, many of which remain consistent with current observational constraints. However, the question of whether and under what physical conditions ABHs actually form horizons remains open~\cite{Frolov:1998,bambi:2017}.\\
\indent A black hole is intuitively defined as a region of spacetime with extremely strong gravity from which not even light can escape, with the horizon serving as its boundary. The concept of a trapped region provides a local (or quasilocal) characterization: it refers to a region where the outgoing, future-directed null geodesics emanating from a spacelike two-surface of spherical topology have negative expansion~\cite{HawEllis:1973,Faraoni:2015,Frolov:1998}. A physical black hole ($\Phi$BH) is defined by the existence of such a trapped region, with the apparent horizon as its boundary \cite{Frolov:2014}. The apparent horizon corresponds to the outer marginally trapped surface.\\
\indent Any consistent model of gravitational collapse that accounts for black hole formation is expected to produce observable black hole features within a finite time as measured by a distant observer. Hence, $\Phi$BHs are assumed to form at a finite time $t_S$, for such an observer \cite{MMT:2022}. In this framework, $\Phi$BHs model ABHs possessing an apparent horizon.\\
\indent The spacetime geometry is commonly accepted to be regular at the apparent horizon \cite{Frolov:1998,bambi:2017} and this regularity is usually ensured by the finiteness of the Ricci and Kretschmann scalars \cite{Bardeen:2018}. Moreover, under the assumption that semiclassical gravity remains valid at the horizon \cite{PP:2009,BMT:2018}, the combined requirements of regularity and observability uniquely determine the near-horizon geometry of spherically symmetric $\Phi$BHs \cite{BMMT:2019,MMT:2022,Terno:2020}. In particular, it was argued that only two distinct classes of dynamical solutions are viable, labeled by $k=0$ and $k=1$, where the value of $k$ specifies the scaling behavior of the effective EMT components with respect to the metric function, $f$ (see Eqs.~(\ref{eqn:line_element}), (\ref{eqn:tau_defn}) and (\ref{eqn:tau_f_scale})). Both classes of solutions necessarily violate the null energy condition (NEC), which is associated with the finite-time formation of the trapped region \cite{HawEllis:1973}. Moreover, this violation occurs in a way that remains consistent with quantum energy inequalities (QEIs) \cite{Kontou:2020}. Together, they provide a consistent framework for black hole formation \cite{MT:2021,BMMT:2019}. Since gravitational-wave signals and photon-ring structures are highly sensitive to the near-horizon geometry, they offer promising observational tests of $\Phi$BHs \cite{Simovic:2024,Maharana:2025}. In fact, the exotic character of a $\Phi$BH is highlighted by its violation of all Buchdahl conditions, in contrast to horizonless UCOs, thereby pointing toward new physics \cite{Soranidis:2025}. Hence, probing the near-horizon structure of $\Phi$BHs becomes essential, both to clarify whether ABHs truly form horizons and thereafter, to explore the possibility of previously unknown physical effects.\\
\indent While all known black hole models belong to either $k=0$ or $k=1$ class, the exclusion of $k<0$ is not completely justified~\cite{Finite_energy:2019,BMMT:2019,MT:2021,MMT:2022,Terno:2020}. In fact, the arguments of Ref.~\cite{Terno:2020} for excluding the $k<0$ solutions are circular as they are essentially based on the assumption that the description of geometry is regular at the apparent horizon in advanced null coordinates. We address this concern by rigorously demonstrating that dynamical solutions with $k <0$ are not consistent $\Phi$BH solutions. We also constrain the near-horizon features of static $\Phi$BHs, which are considered to be the static limit of dynamical $\Phi$BHs.\\
\indent The paper is organized as follows: we provide a brief overview of $\Phi$BHs in semiclassical gravity in section \ref{sec_II}. In section \ref{sec_III}, we analyze the semiclassical Einsteins equations and determine the general form of the near-horizon expansions of the metric functions. In section \ref{sec_IIIa} and \ref{sec_IIIb}, we obtain further constraints on the near-horizon form of the metric functions, and examine possibility of horizon crossing without firewall singularities (which are characterized by divergent proper energy density) at the horizon for the dynamical and static $\Phi$BH solutions respectively. In section \ref{sec_IV}, we conclude the article with a summary of the main results, and a discussion of the limitations of this work. We use $-+++$ signature and the units $\hslash=G=c=1$ throughout the article. And we restrict our analysis only to the spherically symmetric $\Phi$BHs.



\section{Spherically symmetric Physical black holes} \label{sec_II}
Consider a general spherically symmetric spacetime described in Schwarzschild coordinates by the line element
\begin{align}
    ds^2 = -e^{2h(t,r)}f(t,r)\,dt^2 + \frac{1}{f(t,r)}dr^2 + r^2 d\Omega_2\,\,, \label{eqn:line_element}
\end{align}
where \( t \) is the Schwarzschild time and \( r \) is the areal radius. In advanced null coordinates $(v,r)$, the line element takes the form
\begin{align}
    ds^2=-e^{2h_+(v,r)}f_+(v,r)dv^2+2e^{h_+(v,r)}dv\,dr+r^2 d\Omega_2\,\,,
\end{align}
where
\begin{align}
    e^{h_+}dv= e^{h}dt+\frac{1}{f}dr\,\,.
\end{align}
The functions, $f(t,r)$ and $f_+(v,r)$, are coordinate-independent, i.e., $f(t,r)=f_+(v,r)$. Consequently, it enables an invariant definition of the Misner--Sharp mass \cite{Blau:lect_notes,Faraoni:2017} via
\begin{align}
    &C(t, r) = r\left(1 - g^{\mu\nu} \partial_\mu r \partial_\nu r\right)=C_+(v,r) .
\end{align}
\indent In the context of dynamical horizons, the outer marginally trapped surface is defined by \( r = r_g(t) \) (or equivalently, $r=r_+(v)$), where both the functions, $r_g(t)$ and $r_+(v)$, are continuous. It is identified by analyzing the expansion of ingoing and outgoing radial null geodesics. This corresponds to the largest root of the equation \( f(t, r) = 0\) (or $f_+(v,r)=0$) \cite{Faraoni:2017}. \\
\indent Within the semiclassical gravity framework, classical geometric notions, such as trajectories and horizons, remain well-defined \cite{PP:2009,BMT:2018}. The dynamics of spacetime are governed by Einstein's equations, $G_{\mu\nu}=8\pi T_{\mu\nu}$, where the EMT is equal to the expectation value of renormalized EMT, i.e., $T_{\mu\nu}\defeq \langle \hat{T}_{\mu\nu}\rangle_\omega$. In the self-consistent approach, that we use here, no specific assumptions are made a priori regarding the nature of the matter content or the quantum state, $\omega$. The resulting EMT encompasses both the collapsing matter and the associated quantum excitations \cite{MMT:2022}.\\
\indent For convenience, we express Einstein's equations in terms of the metric functions $f$ and $h$ in $(t,r)$ coordinates, in contrast to Refs.~\cite{BMMT:2019,Terno:2020}, which use $C$ and $h$:
\begin{subequations} \label{eqn:Einstein_eqns}
\begin{align}
    \partial_r f &= \frac{1}{r}\left(1 - f - \frac{8\pi r^2 \tau_t}{f}\right)\,, \label{eqn:Einstein_1} \\
    \partial_r h &= \frac{4\pi r(\tau_t + \tau^r)}{f^2}\,,\label{eqn:Einstein_3}\\
    \partial_t f &= -8\pi r e^h \tau_t^r\,, \label{eqn:Einstein_2} 
\end{align}
\end{subequations}
where the effective EMT components, introduced to simplify the subsequent calculations, are defined as \cite{BMMT:2019,Terno:2020,MMT:2022}
\begin{align}
    \tau_t \vcentcolon = e^{-2h} T_{tt}\,\,,\,\,\,\,\tau^r\vcentcolon =T^{rr}\,\,,\,\,\,\,\tau_{t}^{\,\,r}\vcentcolon = e^{-h}T_{t}^{\,\,r}\,\,.\label{eqn:tau_defn}
\end{align}
\indent Similarly, the Einstein's equations in $(v,r)$ coordinates are given by
\begin{subequations}
\begin{align}
    &e^{-h_+}\partial_vC_++f\,\partial_rC_+=8\pi r^2 \theta_v\,\,,\label{eqn:Einstein_1_vr}\\
    & \partial_rC_+=-8\pi r^2\theta_{vr}\,\,,\label{eqn:Einstein_2_vr}\\
    & \partial_rh_+=4\pi r\theta_r\,\,.\label{eqn:Einstein_3_vr}
\end{align}
\end{subequations}
The effective EMT components in $(v,r)$ coordinates are related to those in Eq.~(\ref{eqn:tau_defn}) as
\begin{subequations}
\label{eqn:EMT_vr}
\begin{align}
    &\theta_v\defeq e^{-2h_+}\Theta_{vv}=\tau_t\,\,,\\
    &\theta_{vr}\defeq e^{-h_+}\Theta_{vr}=\frac{\tau_{t}^{\,r}-\tau_t}{f}\,\,,\\
    &\theta_r\defeq \Theta_{rr}=\frac{\tau_t+\tau^r-2\tau_{t}^{\,r}}{f^2}\,\,,
\end{align}
\end{subequations}
where $\Theta_{\mu\nu}$ are the EMT components in the advanced null coordinate system.\\
\indent We can consider the two EMT scalars, $T^{\mu}_{\,\,\mu}$ and $T^{\mu\nu}T_{\mu\nu}$, which are related to the curvature scalars via the semiclassical Einstein equations:

\label{eqn:curv_scalars}
\begin{align}
    T^{\mu}_{\,\,\mu}=-\frac{R^{\mu}_{\,\mu}}{8\pi}\,\,,\,\,\,\, T^{\mu\nu}T_{\mu\nu}= \frac{R_{\mu\nu}R^{\mu\nu}}{64\pi^2}\,\,.
\end{align}
These EMT scalars are related to the EMT components as follows
\begin{subequations}
    \begin{align}
    &T^{\mu}_{\,\,\mu}=\mathrm{T}+2T^{\theta}_{\,\,\theta}\,\,,\\
    &T^{\mu\nu}T_{\mu\nu}=\mathfrak{T}+2(T^{\theta}_{\,\,\theta})^2\,\,,\label{eqn:EMT_scalar_2}
\end{align}
\end{subequations}
where
\begin{align}
    \mathrm{T}\vcentcolon = \frac{\tau^r-\tau_t}{f}\,\,,\,\,\,\,\mathfrak{T}\vcentcolon =\frac{(\tau_{t})^2+(\tau^{r})^2-2(\tau_{t}^{\,\,r})^2}{f^2}\,\,, \label{eqn:EMT_scalars}
\end{align}
and $T^{\theta}_{\,\,\theta}\equiv T^{\phi}_{\,\,\phi}$. Regularity at the apparent horizon is imposed in its weakest form by requiring the curvature scalars $R^{\mu}_{\,\mu}$ and $R^{\mu\nu}R_{\mu\nu}$ to remain finite. This minimal condition ensures the absence of curvature singularities, though it does not guarantee the regularity of all geometric or physical quantities \cite{HawEllis:1973,Joshi,MMT:2022}. It was shown in Ref.~\cite{Terno:2020}, the regularity requirement along with consistency with the semiclassical Einstein's equations, implies that $T^{\theta}_{\,\,\theta}$ diverges weaker than that of $\tau_{a}/f$ where the index $a=\{{}_{t},{}^{r},{}_{t}^{\,\,r}\}$. Building on this result, we demonstrate in Appendix \ref{app2} that each of $\mathrm{T}$, $\mathfrak{T}$, and $T_{\,\,\theta}^{\theta}$ remain finite at the horizon.\\
\indent In a $\Phi$BH spacetime, the apparent horizon is accessible to a distant observer. Consequently, an ingoing null particle should require only a finite coordinate time, as measured by the distant observer, to cross the horizon. This translates into the condition \cite{MMT:2022,Terno:2020}
\begin{align}
\frac{dt}{dr}=-\frac{1}{e^{h}f}
\quad \Rightarrow \quad
t=\int_{r_g}^{r_0} \frac{dr}{e^{h}f} < \infty \,\,,
\label{eqn:obs_condn}
\end{align}
where $r_0>r_g$ and the apparent horizon evolves sufficiently smoothly with respect to the coordinate time $t$.\\
\indent Under the assumption that the effective EMTs scale as powers of $f$ near the horizon of the form
\begin{align}
    \tau_t\sim f^{k_{t}^0}\,\,,\,\,\,\,\tau^r\sim f^{k_{r}^0}\,\,,\,\,\,\,\tau_{t}^{\,\,r}\sim f^{k_{tr}^0}\,\,,\label{eqn:tau_f_scale}
\end{align}
it has been argued in Refs.~\cite{BMMT:2019,MMT:2022,Terno:2020} that only $k_a=0$ and $k_a=1$ $\forall a$ render viable dynamical solutions. However, $k<0$ dynamical solutions were excluded on the assumption that $\theta_v=\tau_t$ needs to be regular at the apparent horizon \cite{Terno:2020}, instead of requiring the regularity of an invariant quantity that is physical.\\
\indent While the finite-time formation of a trapped region in spherically-symmetric collapse necessarily involves violation of the NEC in the vicinity of the apparent horizon \cite{HawEllis:1973}, it must nonetheless respect time-averaged energy conditions \cite{BMMT:2019}. In particular, the time-averaged proper energy density remains finite, albeit negative, in the vicinity of the horizon \cite{Kontou:2020,Kontou:2015}. Moreover, a positively divergent time-averaged proper energy density inevitably leads to geodesic incompleteness at the horizon, thereby identifying the horizon as a surface with matter singularity \cite{HawEllis:1973,Joshi,Ellis:1974}.
\section{Near-horizon features of $\mathbf{\Phi}$BH\lowercase{s}} \label{sec_III}
Assuming a power series expansion of the effective EMTs in terms of $f$ enable a power series expansion in the gap (radial coordinate difference), $x(t,r)\vcentcolon = r - r_g(t)$ \cite{Terno:2020,BMMT:2019}. Since Eq.~(\ref{eqn:Einstein_1}) can be considered to a closed equation for the metric function $f$, it can be readily solved to obtain the leading order term in the power series expansion of $x$. Let us assume that 
\begin{subequations}
    \label{eqn:tau_gen_series}
    \begin{align}
        &\tau_t = \Xi_{t}^{(0)}(t) f^{k_{0}^t}(t,r)+\Xi_{t}^{(1)}(t) f^{k^{t}_1}(t,r)+\,...\,\,,\\
        &\tau^r = \Xi_{r}^{(0)}(t) f^{k_{0}^{r}}(t,r)+\Xi_{r}^{(1)}(t) f^{k^{r}_1}(t,r)+\,...\,\,,\\
        &\tau^{\,r}_{t} = \Xi_{tr}^{(0)}(t) f^{k^{tr}_{0}}(t,r)+\Xi_{tr}^{(1)}(t) f^{k^{tr}_1}(t,r)+\,...\,\,,
    \end{align}
\end{subequations}
where $0<k^{a}_0<k^{a}_1<\,...\,$ for $a=\{t,r,tr\}$. For notational convenience, we are going to denote $k_{0}^t$ as $k$ and $k_{i}^t$ as $k_i$ for $i\geq 1$.\\
\indent The value of the exponents in Eq.~(\ref{eqn:tau_gen_series}) determines the near-horizon form of the metric functions, $f$ and $h$. In semiclassical gravity, the metric functions of $\Phi$BHs admit the following form (see Appendix~\ref{app1})
\begin{subequations}
\label{eqn:f_h_gen_form}
\begin{align}
    &f(t,r)=f_0(t) x^\alpha(t,r) + (\text{higher order terms})\,\,,\label{eqn:f_form}\\
    &h(t,r)=-\delta\ln\frac{x(t,r)}{\xi(t)}+ (\text{higher order terms})\,\,,\label{eqn:h_form}
\end{align}
\end{subequations}
where $\alpha,\delta>0$ are constants. The leading exponent, $\alpha$, is dependent on $k$, $k_1$ as follows -
\begin{enumerate}
    \item $k<1$: $\alpha=\frac{1}{2-k}$,
    \item $k>1$: $\alpha=1$,
    \item $k=1$: $\alpha=\frac{1}{2-k_1}$ and $k_1=(1,2)$.
\end{enumerate}
\hfill \break
\indent The Hadamard condition ensures that quantum fields remain regular at the horizon, as the local behavior in this region must approximate that of the Minkowski vacuum—a direct consequence of the equivalence principle. Consequently, no exotic high-energy phenomena should be observed at the horizon by a freely falling observer \cite{MMT:2022,Almheiri:2013}.

\subsection{Dynamical $\mathbf{\Phi}$BHs}\label{sec_IIIa}

The leading order divergence of the curvature scalars, $R^{\mu}_{\,\,\mu}$ and $R^{\mu\nu}R_{\mu\nu}$, for the metric functions of the form given in Eq.~(\ref{eqn:f_h_gen_form}), are obtained to be
\begin{widetext}
\begin{subequations}
    \begin{align}
    &R^{\mu}_{\,\,\mu}=\frac{1}{f_0 \xi^{2\delta}}\left[(\alpha-2\delta)(1-\alpha+\delta)f_{0}^{2}\xi^{2\delta}+\alpha(1+\alpha-\delta)r_{g}'^2x^{2(\delta-\alpha)}\right] x^{\alpha-2}+(\text{higher order terms})\,\,,\\
    & R^{\mu\nu}R_{\mu\nu} = \frac{1}{4f_{0}^{2} \xi^{4\delta}}\left[(\alpha-2\delta)(1-\alpha+\delta)f_{0}^{2}\xi^{2\delta}+\alpha(1+\alpha-\delta)r_{g}'^2x^{2(\delta-\alpha)}\right]^2 x^{2\alpha-4}+(\text{higher order terms})\,\,,
\end{align}
\end{subequations}
\end{widetext}
where $r_{g}'=\frac{dr_g}{dt}$.\\
\indent These divergences can only be removed by the conditions
\begin{subequations}
    \begin{align}
    &\delta=\alpha\,\,,\\
    & r_g' = \pm f_0 \xi^\alpha\,\,,
\end{align}
\end{subequations}
and as a consequence, this form of the metric functions also satisfy the observability condition in Eq.~(\ref{eqn:obs_condn}).\\
\indent Thus, the requirement of the regularity of the curvature scalars gives us the following form of the metric functions
\begin{subequations}
\label{eqn:dyn_metric_fn_expansion}
    \begin{align}
    & f(t,r) = f_0(t)x^\alpha(t,r) (1+f_1(t)x^{\beta_1}(t,r)+...)\,\,,\label{eqn:f_expr_dyn}\\
    & h(t,r) = - \alpha \ln \frac{x(t,r)}{\xi(t)}+h_1(t)x^{\beta_1}(t,r)+...\,\,.
\end{align}
\end{subequations}
From Eq.~(\ref{eqn:dyn_metric_fn_expansion}), the combination, $e^hf$, admits the following expansion
\begin{align}
    e^hf=f_0\xi^\alpha\left(1+\mathfrak{F}\,x^{\chi}\right)\,\,,\label{eqn:ehf_sublead}
\end{align}
where $\chi=\beta_n>0$ for some $n\in \mathbb{N}$ and $\mathfrak{F}\neq 0$ is a polynomial function of the metric function coefficients, $\{f_i\}_{i\geq 1}$ and $\{h_i\}_{i\geq 1}$.\\
\indent Since $r=r_g(t)$ is a timelike hypersurface, a timelike test particle can travel along the apparent horizon, satisfying the relation \cite{BMMT:2019, MMT:2022}
\begin{align}
    & d\tau^2=e^{2h}f \,dt^2-\frac{1}{f}(r_{g}')^2 dt^2\notag\\
    \Rightarrow \,\,& \left(\frac{dt}{d\tau}\right)^2\approx\frac{1}{2f_0\xi^{2\alpha}\mathfrak{F}}x^{\alpha-\chi}\,\,\notag\\
    \Rightarrow \,\,& \mathfrak{F}>0\,\,.\label{eqn:F_sub_gtr_0}
\end{align}
Applying this result to Eq.~(\ref{eqn:ehf_sublead}) further implies that
\begin{align}
    \partial_r(e^hf)>0\,\,.
\end{align}
Moreover, Eqs.~(\ref{eqn:Einstein_1}) and (\ref{eqn:Einstein_3}) render the relation
\begin{align}
    \partial_r(e^hf)=\frac{e^h}{r}\left[1-f+4\pi r^2\mathrm{T}\right]\,\,. \label{eqn:eh_f_eqn}
\end{align}
\indent The case $r_g'<0$ corresponds to evaporating $\Phi$BH solutions, while $r_g'>0$ describes their white hole counterparts \cite{MMT:2022}. Considering the $\Phi$BH case, we take
\begin{align}
    r_g'=-f_0\xi^\alpha\,\,.
\end{align}
Since $k\geq 0$ dynamical solutions has been satisfactorily dealt with in Refs. \cite{BMMT:2019,MMT:2022,Terno:2020}, we focus our attention on the $k<0$ solutions, and consequently, from Table \ref{tab:leading_semiclassical}, we have $\alpha=\frac{1}{2-k}<\frac{1}{2}$. The finiteness of the EMT quantities, $\mathrm{T}$ and $\mathfrak{T}$, at the apparent horizon (see Appendix \ref{app2}) allows us to express $\tau^{r}$ and $\tau_{t}^{\,r}$ in terms of $\tau_t$, $\mathrm{T}$ and $\mathfrak{T}$ as 
\begin{subequations}
\label{eqn:tau_reln_ttheta}
\begin{align}
    & \tau^r=\tau_t+\mathrm{T}f\,\,,\\
    &\tau_{t}^{\,r}=\left[\frac{(\tau_{t})^2+(\tau^{r})^2-\mathfrak{T}f^2}{2}\right]^{1/2}\notag\\
    \Rightarrow\, & \tau_{t}^{\,r}\approx \tau_t+\frac{\mathrm{T}f}{2}+\frac{1}{8\tau_t}(\mathrm{T}^2-2\mathfrak{T})f^2\,\,,
\end{align}
\end{subequations}
and using relations in Eqs.~(\ref{eqn:EMT_vr}) and (\ref{eqn:EMT_scalars}), we obtain 
\begin{align}
    \theta_{vr} \sim\mathcal{O}\left(f^0\right)\,\,,\,\,\,\,\theta_r\sim\mathcal{O}\left(f^{\epsilon}\right)\,\,,
\end{align}
where $\epsilon \geq |k|$. From Eqs.~(\ref{eqn:Einstein_2_vr}) and (\ref{eqn:Einstein_3_vr}),
\begin{align}
    &e^{h_+}\sim \mathcal{O}\left(f^0\right)\,\,,\\
    & C_+\approx r_+(v)-8\pi r_{+}^2(v)\theta_{vr}|_{r=r_+(v)}(r-r_+(v))\,\,,\label{eqn:C_plus_exp}
\end{align}
and substituting these in Eq.~(\ref{eqn:Einstein_1_vr}) gives
\begin{align}
    r_+'(v)\sim  \mathcal{O}\left(f^{-|k|}\right)\,\,.
\end{align}
This essentially implies that $r_+'(v)$ is undefined for all values of $v$ contradicting Eq.~(\ref{eqn:C_plus_exp}). Physically, this means that black hole evaporation is undefined for $k<0$ solutions, and hence, there are no $\Phi$BH solutions in semiclassical gravity where the effective components scale as $f^k$ with $k<0$.

\subsection{Static $\mathbf{\Phi}$BHs}\label{sec_IIIb}

While static solutions serve as useful idealizations, ABHs are generally embedded in environments dominated by high-energy processes and are therefore not expected to be strictly static. Nevertheless, the static case remains of interest: it can serve as a transient phase between fully dynamical models or as a quasistatic approximation of a slowly evolving black hole. Moreover, characteristic features such as photon rings and QNM frequencies, are more tractable to analyze in the presence of a static limit \cite{Mishra:2019}.

Among static configurations, those with $k<1$ are excluded as viable $\Phi$BH candidates because the curvature scalar $R^{\mu\nu}R_{\mu\nu}$ (cf. Eqs.~(\ref{eqn:EMT_scalar_2}) and (\ref{eqn:EMT_scalars})) cannot remain regular. Typical static black hole models correspond to $k=1$ solutions and the infall time for an ingoing null particle is infinite according to a distant observer~\cite{BMMT:2019,Terno:2020,MMT:2022,MT:2021}. Here, we show that the latter is not necessarily true. Furthermore, certain cases admit NEC violation consistent with the QEIs.

Since $\partial_t$ is the timelike Killing vector field in static spacetime, the metric functions are independent of $t$ \cite{HawEllis:1973}. The curvature scalars, $R^{\mu}_{\,\,\mu}$ and $R^{\mu\nu}R_{\mu\nu}$, for the metric functions in Eq.~(\ref{eqn:f_h_gen_form}) admit a Laurent series and the leading order divergence are obtained to be
\begin{subequations}
\label{eqn:Rs_stat}
\begin{align}
    &R^{\mu}_{\,\,\mu}=f_0(\alpha-2\delta)(1-\alpha+\delta) x^{\alpha-2}\notag\\
    &\,\,\,\,\,\,\,\,\,\,\,\,\,\,\,\,\,\,\,\,\,\,\,\,\,\,\,\,\,\,\,\,\,\,\,\,\,\,\,\,\,\,\,\,\,\,\,\,+(\text{higher order terms})\,\,,\label{eqn:Rs1_stat}\\
    & R^{\mu\nu}R_{\mu\nu} = \frac{f_{0}^2}{4}(\alpha-2\delta)^2(1-\alpha+\delta)^2x^{2\alpha-4}\notag\\
    &\,\,\,\,\,\,\,\,\,\,\,\,\,\,\,\,\,\,\,\,\,\,\,\,\,\,\,\,\,\,\,\,\,\,\,\,\,\,\,\,\,\,\,\,\,\,\,\,+(\text{higher order terms})\,\,.\label{eqn:Rs2_stat}
\end{align}
\end{subequations}
This divergence can be eliminated if the exponents are related by $\delta=\frac{\alpha}{2}$ or $\delta=\alpha-1$. Only the former condition
\begin{align}
    \delta=\frac{\alpha}{2}\,\,,\label{eqn:alpha_delta_stat_reln}
\end{align}
is compatible with both the regularity and the observability conditions in Eq.~(\ref{eqn:Rs_stat}) and Eq.~(\ref{eqn:obs_condn}) respectively. Note that we cannot have the metric function $h$ to be regular at the horizon i.e., $\delta=0$, since the regularity condition in Eq.~(\ref{eqn:alpha_delta_stat_reln}) implies $\alpha=1$, which in turn violates the observability condition in Eq.~(\ref{eqn:obs_condn}).\\
\indent Along with addressing curvature singularities signaled by divergent scalar invariants, it is also essential to examine possible matter singularities \cite{HawEllis:1973,Ellis:1974}. To this end, we first determine whether radially infalling timelike test particles cross the horizon in finite proper time, and then investigate whether the integrated proper energy density measured along their worldlines diverges at the horizon.\\
\indent The four-velocity of a radially moving timelike observer, $u^\alpha=\left(\dot{t},\dot{r},0,0\right)$, parametrized by the proper time, $\tau$, satisfies the normalization condition
\begin{align}
    -e^{2h}f\dot{t}^2+\frac{1}{f}\dot{r}^2=-1
    \Rightarrow\,\, \dot{t}=\frac{\sqrt{\dot{r}^2+f}}{e^hf}\,\,,\label{eqn:norm_condn}
\end{align}
where $\dot{t}\equiv\frac{dt}{d\tau}$ and $\dot{r}\equiv\frac{dr}{d\tau}$.\\
\indent In a regular spacetime, a timelike particle cannot become null within a finite proper time \cite{HawEllis:1973}, and thus, its four acceleration, $a^\mu \defeq u^\nu u^{\mu}_{\,\,;\nu}$, remains finite, i.e.,
\begin{align}
    |a^\mu a_\mu|<\infty\,\,.\label{eqn:acc_finite}
\end{align}
For a radially moving timelike observer, the normalization condition in Eq.~(\ref{eqn:norm_condn}) implies that the four-acceleration components are related as
\begin{align}
    a^t=\frac{\dot{r}}{e^hf\sqrt{\dot{r}^2+f}}\,a^{r}\,\,,
\end{align}
so that Eq.~(\ref{eqn:acc_finite}) reduces to
\begin{align}
    &\frac{(a_{r})^2}{\dot{r}^2+f}<\infty \notag\\
    \Rightarrow \,&\ddot{r}+\frac{1}{2}\partial_rf+\left(\dot{r}^2+f\right)\partial_rh\sim\mathcal{O}\left(\sqrt{\dot{r}^2+f}\right)\,\,.\label{eqn:geo_static}
\end{align}
To leading order, for a radially infalling observer, Eq.~(\ref{eqn:geo_static}) simplifies to 
\begin{align}
    &\ddot{x}-\frac{\alpha}{2x}\dot{x}^2\approx 0\,\,,
\end{align}
and in this case, the above equation can be solved as 
\begin{align}
    &\dot{x}=-Kx^{\alpha/2} \notag\\
    \Rightarrow\,\,&x^{1-\frac{\alpha}{2}}=K\left(1-\frac{\alpha}{2}\right)(\tau_c-\tau)\,\,,\notag\\
    &t-t_c=\frac{1}{f_0}\left(\frac{K^2+f_0}{\xi^\alpha}\right)^{1/2}(\tau-\tau_c),
\end{align}
where $K>0$ is a constant. Thus, the particle takes a finite coordinate time, $t_c$, and finite proper time, $\tau_c$, to cross the horizon. \\
\indent  Turning now to physical observables, we use Eq. (\ref{eqn:alpha_delta_stat_reln}) and Table~\ref{tab:leading_semiclassical} to compute the proper energy density, $\rho_A$, as measured by a comoving observer Alice:
\begin{align}
    \rho_A&=T_{\mu\nu}u^\mu u^\nu\notag\\
    & =\begin{cases}
    \Xi_{t}^{(0)}-\frac{K^2}{8\pi r_{g}}, & \text{if $k=1$ and $\alpha=1$},\\
    \frac{1}{8\pi r_{g}^2}, & \text{if $k=1$ and $1<\alpha<2$},\\
    -\frac{K^2}{8\pi r_{g}^2}, & \text{if $k>1$}.
  \end{cases}
\end{align}
\indent For $k=1$ with $\alpha=1$, the proper energy density at the horizon remains finite, and sufficiently fast moving observers can register NEC violation. For $k=1$ with $1<\alpha<2$, the energy density at the horizon is finite, strictly positive, and is also equal to the corresponding value for the dynamical $k=1$ solution \cite{MMT:2022,MT:2021}. In contrast, $k>1$ solutions exhibit NEC violation near the horizon consistent with the QEIs. However, the NEC violation in the static case does not necessarily imply that a trapped region can form within finite time. Since the formation of a trapped region is inherently dynamical, a static $\Phi$BH can only be understood as the limiting case of an underlying dynamical $\Phi$BH.\\
\section{Discussion} \label{sec_IV}
We completed the classification of dynamical spherically-symmetric $\Phi$BH solutions whose effective EMT in the near-horizon limit behave as $f^k$, with $k< 0$. These solutions cannot be excluded by the arguments provided in Ref.~\cite{Terno:2020}. We showed that all such solutions that satisfy the conditions of finite time of formation and finiteness of the curvature scalars at the apparent horizon do not admit a consistent black hole evaporation scenario in semiclassical gravity.\\
\indent Furthermore, we have introduced a new class of static $\Phi$BH solutions which are compatible with the observability and regularity requirements. In particular, static solutions with $k<1$ are not viable. For $k=1$, the proper energy density is finite, with NEC violation occurring in only one class. Solutions with $k>1$ exhibit NEC violation, in agreement with the QEIs.\\
\indent The analysis presented here is easily extendible to the white hole counterparts of $\Phi$BHs since an analogous calculation in the retarded null coordinates shows the inconsistency for $k<0$ solutions. Precisely speaking, the evolution of the anti-trapping horizon as a function of the retarded null coordinate is ill-defined for $k<0$ white hole solutions. Although we find static $\Phi$BH solutions that exhibit NEC violation consistent with the QEIs, their physical relevance can only be assessed if they arise as the static limit of a consistent dynamical $\Phi$BH solution. This issue is under current investigation and will be addressed in future work. 
\hfill \break
\section{Acknowledgements}

We thank Daniel Terno for valuable critical comments and Ioannis Soranidis for helpful discussions. SM and RV are supported by an International Macquarie University Research Excellence Scholarship.\\



\appendix

\section{Leading-order terms of the metric functions} \label{app1}
The near-horizon behaviour of the metric function, $f$, can be obtained by solving Eq.~(\ref{eqn:Einstein_1}) and the resulting solutions are classified in Table~\ref{tab:leading_semiclassical}.\\
\indent We can use Eq.~(\ref{eqn:Einstein_3}) to determine $h$, which is dependent on the form of $\tau_t$ and $\tau^r$. Let us assume
\begin{align}
    \tau_t + \tau^r \equiv  \Xi_{t+r}^{(0)} f^{k'}\,\,,
\end{align}
where 
\begin{align}
    k'
    \begin{cases}
        &=\text{min}(k,k_{0}^r)\,\,,\,\,\,\,\text{if $k_{0}^{r}\neq k$ ,}\\
        &=k\,\,,\,\,\,\,\text{if $k_{0}^{r}=k$ and $\Xi_{t}^{(0)}+\Xi_{r}^{(0)}\neq 0$ ,}\\
        &> k\,\,,\,\,\,\,\text{if $k_{0}^{r}=k$ and $\Xi_{t}^{(0)}+\Xi_{r}^{(0)}= 0$ .}\\
    \end{cases}
\end{align}

\begin{widetext}

\renewcommand{\arraystretch}{1.5}  

\begin{table}[ht]
\centering
\begin{tabular}{ |c|c|c|c| } 
\hline
$k$ & Form of the effective EMTs  & $f$ & $h$ \\
\hline
\hline
$k<1$ & $\Xi_{t}^{(0)}<0$, $k'=k$ & $\left[8\pi r_g |\Xi_{t}^{(0)}|(2-k)x\right]^{\frac{1}{2-k}} + ...$ & \rule{0pt}{4ex}$\frac{\Xi_{t+r}^{(0)}}{2|\Xi_{t}^{(0)}|(2-k)}\ln\frac{x}{\xi}+...$ \\ [2ex]
\hline
\hline
\multirow{2}{2.5em}{$k=1$} & \makecell[c]{$\Xi_{t}^{(0)} < \frac{1}{8\pi r_{g}^{2}}=-\Xi_{r}^{(0)}$,\\ $k'= 1$} & $\frac{(1-8\pi r_{g}^{2}\Xi_{t}^{(0)})}{r_g}x + ...$ & \rule{0pt}{5ex} $\frac{\Xi_{t+r}^{(0)}}{2\left(\frac{1}{8\pi r_{g}^{2}}-\Xi_{t}^{(0)}\right)}\ln\frac{x}{\xi}+...$ \\ [4ex]
\cline{2-4}
& \makecell[c]{$\Xi_{t}^{(0)} = \frac{1}{8\pi r_g^2} = -\Xi_{r}^{(0)}$ \\ 
$1<k'=k_1<2$, \, $\Xi_{t}^{(1)}<0$} & $\left[8\pi r_g |\Xi_{t}^{(1)}|(2-k_1)x\right]^{\frac{1}{2-k_1}}+ ...$ & \rule{0pt}{5ex} $\frac{\Xi_{t+r}^{(0)}}{2|\Xi_{t}^{(1)}|(2-k_1)}\ln\frac{x}{\xi}+....$ \\ [3ex]
\cline{2-4}
& \makecell[c]{$\Xi_{t}^{(0)} = \frac{1}{8\pi r_{g}^{2}}$, $k_1=2$, \\ $\Xi_{t}^{(1)}<0$} & $-\frac{x^2}{r_{g}^2}+ ...$ & \rule{0pt}{5ex} --- \\ [3ex]
\hline
$k>1$ & $k'=1$ & $\frac{x}{r_g} + ...$ & \rule{0pt}{2ex} $4\pi r_{g}^{2}\Xi_{r}^{(0)}\ln\frac{x}{\xi}+...$ \\ [1ex]
\hline
\end{tabular}
\caption{Leading order terms of the metric functions, $f$ and $h$, depending on the value of the effective EMT parameters, $k$, $k_1$, $k_r$, $\Xi_{t}^{(0)}$, $\Xi_{t}^{(1)}$, $\Xi_{r}^{(0)}$, $\Xi_{r}^{(1)}$, and $\Xi_{t+r}^{(0)}$. Constraints on these parameters are derived by requiring real-valued power series solutions from the Eqs. (\ref{eqn:Einstein_eqns}) and ensuring the regularity of the curvature scalars.}
\label{tab:leading_semiclassical}
\end{table}

\end{widetext}
We explicitly solve it for $k<1$ and show that the metric function, $h$, is of the form in Eq.~(\ref{eqn:h_form}). The calculations for the other cases follow similarly.
From Eq.~(\ref{eqn:Einstein_3}), we get
\begin{align}
    \partial_r h & \approx 4\pi r \Xi^{(0)}_{t+r} f^{k'-2}\notag\\
    \Rightarrow h&\approx \begin{cases}
\frac{4\pi r(2-k)\left[8\pi|\Xi_t^{(0)}|r_g(2-k)\right]^{-\frac{2-k'}{2-k}}x^{\frac{k'-k}{2-k}}}{k'-k}  \,\,,\,\,\,\,k'\neq k\\
\frac{\Xi_{t+r}^{(0)}}{2|\Xi_{t}^{(0)}|(2-k)}\ln\frac{x}{\xi}\,\,,\,\,\,\,k'= k\,\,.
\end{cases} 
\label{eqn:h_gen_form}
\end{align}
In the latter case in Eq.~(\ref{eqn:h_gen_form}), the leading order divergence in the scalar curvatures can be eliminated, as demonstrated in Eq.~(\ref{eqn:Rs_stat}). This cancellation, however, does not occur in the former case.\\

\section{Finiteness of $\mathrm{T}$, $\mathfrak{T}$ and $T_{\theta\theta}$ at the apparent horizon for $k<1$ dynamical $\Phi$BH solutions}
\label{app2}
Consider the effective EMT components of the form
\begin{subequations}
\begin{align}
    &\tau_{t}\approx\Xi_{t}^{(0)}f^k+\Xi_{t}^{(1)}f^{k_1}+\mathcal{O}(f)\,\,,\\
    &\tau^{r}\approx\Xi_{r}^{(0)}f^k+\Xi_{r}^{(1)}f^{k_1}+\mathcal{O}(f)\,\,,\\
    &\tau_{t}^{r}\approx\Xi_{tr}^{(0)}f^k+\Xi_{tr}^{(1)}f^{k_1}+\mathcal{O}(f)\,\,,
\end{align}
\end{subequations}
where $k<1$ (see Table \ref{tab:leading_semiclassical}). Requiring regularity of the curvature scalars and consistency with Einstein’s equations constrains the leading-order coefficients as \cite{Terno:2020}
\begin{align}    
    \Xi_{t}^{(0)}=\Xi_{r}^{(0)}=\Xi_{tr}^{(0)}\,\,.
\end{align}
Furthermore, imposing regularity of the curvature scalars also yields a relation between the subleading-order coefficients
\begin{align}    \Xi_{t}^{(1)}=\Xi_{r}^{(1)}=\pm\Xi_{tr}^{(1)}\,\,.\label{eqn:EMT_coeff_reln}
\end{align}
\indent Next, consider the following linear combination of Eq.~(\ref{eqn:Einstein_1}) and Eq.~(\ref{eqn:Einstein_2}) which gives us the relation
\begin{align}
    &\left(\partial_r-\frac{\partial_r}{e^hf}\right)f=\frac{8\pi r(\tau_{t}^{\,\,r}-\tau_t)}{f}+\mathcal{O}(f^0)\label{eqn:f_comb_eqn}\\
    \Rightarrow &\left(\partial_r-\frac{\partial_t}{f_0\xi^\alpha}\right)f + \frac{\mathfrak{F}x^\chi}{f_0\xi^\alpha}\partial_tf\approx\frac{8\pi r(\tau_{t}^{\,\,r}-\tau_t)}{f}\,\,. \label{eqn:f_consistency}
\end{align}
From Eq.~(\ref{eqn:Einstein_1}), $f$ admits the following power series expansion 
\begin{align}
    f=f_0x^{\frac{1}{2-k}}\left(1+f_1 x^{\frac{k_1-k}{2-k}}+...\right)\,\,,\label{eqn:f_expansion_Ttheta}
\end{align}
which implies the following scaling behaviour of the expressions in Eq.~(\ref{eqn:f_consistency}),
\begin{subequations}
\label{eqn:f_consistency_LHS}
\begin{align}
    &\left(\partial_r-\frac{\partial_t}{f_0\xi^\alpha}\right)f\sim \mathcal{O}\left(x^{\frac{1}{2-k}}\right)\,\,,\\
    & \frac{\mathfrak{F}x^\chi}{f_0\xi^\alpha}\partial_tf\sim \mathcal{O}\left(x^{\chi-\frac{k_1-k}{2-k}-\frac{1-k_1}{2-k}}\right)\,\,,
\end{align}
\end{subequations}
where $\chi>\frac{k_1-k}{2-k}$ from Eq.~(\ref{eqn:eh_f_eqn}) and Eq.~(\ref{eqn:EMT_coeff_reln}). Considering $\Xi_{t}^{(1)}=\Xi_{r}^{(1)}=-\Xi_{tr}^{(1)}$, we obtain
\begin{align}
    \frac{8\pi r(\tau_{t}^{\,\,r}-\tau_t)}{f}\sim \mathcal{O}\left(f^{k_1-1}\right)\sim \mathcal{O}\left(x^{-\frac{1-k_1}{2-k}}\right)\,\,.\label{eqn:f_consistency_RHS}
\end{align}
From Eq.~(\ref{eqn:f_consistency_LHS}) and Eq.~(\ref{eqn:f_consistency_RHS}), we can conclude that  $\Xi_{t}^{(1)}=\Xi_{r}^{(1)}=-\Xi_{tr}^{(1)}$ is not consistent with the Einstein's equation as it violates Eq.~(\ref{eqn:f_consistency}). Therefore, the only consistent choice is
\begin{align}
    \Xi_{t}^{(1)}=\Xi_{r}^{(1)}=\Xi_{tr}^{(1)}\,\,,\label{eqn:tau_sub_coeff_reln}
\end{align}
which allows us to express $\tau^r$ and $\tau_{t}^{\,r}$ in terms of $\tau_t$ as
\begin{subequations}
\label{eqn:tau_relns}
\begin{align}
    &\tau^r = \tau_t + \mathfrak{B}_r f^{k_1}+\mathcal{O}(f)\,\,, \\
    &\tau_{t}^{\,r}=\tau_t + \mathfrak{B}_{tr} f^{k_1}+\mathcal{O}(f)\,\,,
\end{align}
\end{subequations}
where $k_1<1$. Imposing the regularity of the curvature scalars and Eq.~(\ref{eqn:tau_sub_coeff_reln}) then implies
\begin{align}
    \mathfrak{B}_{r}=\mathfrak{B}_{tr}=0\,\,.\label{eqn:B_tau_reln}
\end{align}
Thus, from Eq.~(\ref{eqn:tau_relns}) and Eq.~(\ref{eqn:B_tau_reln})
\begin{align}
    &\tau^r = \tau_t +\mathcal{O}(f)\,\,, \\
    &\tau_{t}^{\,r}=\tau_t + \mathcal{O}(f)\,\,,
\end{align}
which implies that $\mathrm{T}$ and $\mathfrak{T}$, and consequently, $T_{\theta\theta}$, are finite at the apparent horizon.

\bibliographystyle{apsrev4-2}

\newpage{\pagestyle{empty}\cleardoublepage}

\end{document}